\documentclass[pdf]{PoS}
\usepackage{amsmath}

\newcommand{\phibar}{\overline{\phi}}
\newcommand{\psibar}{\overline{\psi}}

\newcommand{\Real}{\mathop{\rm Re}\nolimits}
\newcommand{\tr}{\mathop{\rm tr}\nolimits}

\title{Numerical results of
two-dimensional $N=(2,2)$ super Yang-Mills theory}
\ShortTitle{Numerical results of two-dimensional $N=(2,2)$ super Yang-Mills theory}

\author{Hidenori Fukaya, \speaker{Issaku Kanamori}, Hiroshi Suzuki and Tomohisa Takimi\\
Theoretical Physics Laboratory, RIKEN, 2-1 Hirosawa, Wako, 
Saitama 351-0198, Japan\\
E-mail: \email{hfukaya@riken.jp}, \email{kanamori-i@riken.jp}, \email{hsuzuki@riken.jp}, \email{ttakimi@riken.jp}
}

\abstract{
We report the results of a numerical simulation of a lattice formulation of the
two-dimensional $N=(2,2)$ super Yang-Mills theory proposed by
Suzuki and Taniguchi~\cite{Suzuki-Taniguchi}. 
We measure the 1-point functions and
2-point functions. The scenario is that only tuning of the scalar mass to a
specific value gives a supersymmetric continuum limit. Our results are
consistent with this scenario although conclusive results on the restoration of
supersymmetry have not been obtained.

}

\FullConference{The XXV International Symposium on Lattice Field Theory\\
		 July 30-4 August 2007\\
		 Regensburg, Germany}

\begin{document}

\maketitle

\section{Introduction}
A lattice formulation of the supersymmetric gauge theory is 
important to understand the non-perturbative aspects.
Recently some formulations are proposed by several authors~\cite{Giedt:review}.
Most of the models use the fact that a nilpotent part of 
the supersymmetry can be kept on a lattice for $N\geq 2$ cases 
using the topological twist and the relations among some of them are now
becoming transparent~\cite{relationships, Damgaard:2007xi, Damgaard:2007be}.
There is also an attempt to keep all the
supersymmetry~\cite{Hokkaido,Hokkaido-3d}.
Another approach is a model without exact 
supersymmetry~\cite{Suzuki-Taniguchi} which uses the fact that
in two-dimensional case, because of the super-renormalizability,
there are only a few fine-tuning parameters to obtain the
supersymmetric continuum limit. 
One of the merits of lattice formulation is to enable one to perform numerical
simulations. In two-dimensional case, some results are known
for the super Yang-Mills theory with the topological
twist~\cite{Catterall:simulation,on_computer}.

In this talk, we report the result of a numerical simulation
of the model without exact supersymmetry proposed in \cite{Suzuki-Taniguchi}.
We measure some 1-point and 2-point functions.

\section{Model and Algorithm}

The target theory in the continuum is the 2-dimensional $N=(2,2)$
supersymmetric Yang-Mills theory which is obtained by a dimensional reduction
from the 4-dimensional $N=1$ super Yang-Mills. The lattice action is defined as
a lattice version of a 4-dimensional action on $L \times L \times 1 \times 1$ 
lattice together with a scalar mass counter term $S_{\rm counter}$:\footnote{
For the notational details, see \cite{Suzuki-Taniguchi}.
}
\begin{equation}
 S=S_{\text G} + S_{\text F} + S_{\rm counter}.
\end{equation}
The bosonic part is a plaquette action
\begin{align}
  S_{\rm G}[U]
   &=\frac{\beta}{2N_c}\sum_{x\in\Gamma}\sum_{M,N}
     \Real\tr\bigl\{1-P(x,M,N)\bigr\},
\\
  P(x,M,N)
   &=U(x,M)U(x+a\hat M,N)U(x+a\hat N,M)^{-1}U(x,N)^{-1},
\end{align}
where we use $SU(N_C)$ gauge link variables, $U(x,\mu)=\exp(agA_\mu^a(x)T^a)$,
($\mu=0$, $1$) and compact scalar fields, $U(x,2)=\exp(ag\varphi^a(x)T^a)$ and
$U(x,3)=\exp(ag\phi^a(x)T^a)$. The coupling constant $g$ is related to $\beta$
through $\beta=2N_c/a^2g^2$. The fermion action consists of the Wilson-Dirac
operator
\begin{align}
 S_{\rm F}[U,\lambda]
  &=-a^2\sum_{x\in\Gamma}
   \tr\{\lambda(x)CD_{\text{w}}\lambda(x)\}, &
 D_{\text{w}}
  &=\frac{1}{2}\sum_{M=0}^3\{ \Gamma_M(\nabla_M^* + \nabla_M) - a\nabla_M^*\nabla_M\},
\end{align}
with covariant differences for the adjoint representation $\nabla_M$
\begin{align}
 \nabla_M\lambda(x)
  &=\frac{1}{a}
   \left\{U(x,M)\lambda(x+a\hat M)U(x,M)^{-1}-\lambda(x)\right\}
\end{align}
and its adjoint $\nabla^*_M$\ . The counter term is
\begin{equation}
 S_{\rm counter}[U]
   =-\mathcal{C} N_c\sum_{x\in\Gamma}
   \left( \tr\{U(x,3)+U(x,3)^{-1}-2\} + \tr\{U(x,2)+U(x,2)^{-1}-2\}\right),
\end{equation}
where $\mathcal{C}=0.65948255(8)$.
The counter term is intended to cancel the radiative corrections
to the scalar mass term.  Other corrections which might appear 
in the effective action are suppressed in the continuum limit because
of the super-renormalizability of this theory.
All possible divergences in the sub-diagrams in the perturbative
expansion are suppressed.  
It should be noted, however,  
that it does not guarantee the supersymmetry of composite operators.

In our numerical simulation, we use quenched gauge configurations generated by
using the Hybrid Monte Carlo algorithm. The fermion contribution is introduced
as a reweighting by the Pfaffian. In the continuum limit, the model has a real
and positive Pfaffian. In fact the direct calculation for some sample
configurations shows that the Pfaffian is real and positive in our parameter
region. Therefore we use a positive square root of the determinant which 
numerical cost is
much less expensive. We set a bare fermion mass $m=0$. The lattice size is
$8\times 8$ for 1-point functions and $12 \times 12$ for 2-point functions.
Since the coupling is $\beta=2N_c/a^2g^2$, the continuum limit is the
$\beta \to \infty$ limit. We set $3 \leq \beta \leq 40$.
The gauge group is $SU(2)$.

We summarize the parameters and the numbers of independent configurations in
table~\ref{tab:numbers}.
\TABLE{
%\begin{table}[tb]
 \begin{tabular}{l|c|cccccccc}
 \multicolumn{1}{c|}{$\mathcal{C}$} & $\beta$& 40& 20  & 13  & 10  & 8   & 7   & 5   & 3\\
 \hline
  0.001   &             & 301 & 301 & 301 & 301 & 301 & 301 & -   & - \\
  0.10939 & num.        & -   & -   & -   & -   & 301 & -   & -   & - \\
  0.4     & of          & -   &9801 & -   & -   &9801 & -   & -   & - \\
  0.65948255& configs.  & 801 &9801 & 801 & 801 &9801 & 801 & 801 & 801\\
  1.0     &             & -   &9801 & -   & -   &9801 & -   & -   & - \\
  1.5     &             & -   &9801 & -   & -   &9801 & -   & -   & - \\
 \hline
 \multicolumn{2}{c|}{$ag$} &0.316&0.447&0.555&0.632&0.707&0.756&0.894&1.154 
 \end{tabular}
 \caption{The numbers of configurations for each parameter set on $8\times 8$
lattice. 
% The lattice size is $8\times 8$. 
% The fermion bare mass is $0$ for all cases.
}
 \label{tab:numbers}
%\end{table}
}

\section{One-point functions}

It is of our interest to measure the vacuum expectation value of a
supercharge-exact operator, $\langle Q\mathcal{O}\rangle$, because it must
vanish in the supersymmetric continuum limit. It should be noted, however, that
$\langle Q\mathcal{O}\rangle$ can be non-zero (a finite renormalization),
depending on the definition of the composite operator $Q\mathcal{O}$ which does
not necessarily preserve the supersymmetry.

We make use of a scalar part of the topological twisted supercharges $Q$
and observe $Q$-exact 1-point functions used
in~\cite{Catterall:simulation,on_computer}.
Since we have no exact supersymmetry at finite lattice spacings,
first we write down the continuum relations and then discretize them.
We define the scalar supercharge $Q$ in the continuum as follows:
\begin{align}
 &&
 Q A_\mu^a     &= \psi_\mu^a\ ,  &
 Q \psi_\mu^a  &= iD_\mu \phi'^a,& 
 Q \phi'^a     &= 0\ ,             \\
 &&
 Q \phibar'^a  &= \eta^a,        &
 Q \frac{\eta^a}{2} &= -\frac{i}{2}g f_{abc}\phi'^b \phibar'^c\ , &
 Q \chi^a      &= iF_{01}^a\ . &&
\end{align}
Here we introduce scalar fields $\phi' = \varphi+i\phi$ 
and $\phibar'=\varphi - i\phi$.
Fermions in the twisted basis $\psi_\mu$, $\eta$ and $\chi$ are given by
liner combinations of the components of $\lambda$:
$ ( \eta/2, \chi, \psi_0, \psi_1)^T \equiv T \lambda$.

We use the following three $\mathcal{O}_i$'s:
\begin{align}
 \mathcal{O}_1
      &=-\frac{i}{8}g \eta^a f_{abc}\phi'^b \phibar'^c, &
 \mathcal{O}_2 
      &= -2i\chi^a F_{01}^a\ , &
 \mathcal{O}_3
      &= -\frac{i}{2}\psi_\mu^a D_\mu \phibar'^a.
\end{align}
We divide $Q\mathcal{O}_i$ into two parts $Q\mathcal{O}_i =F_i +B_i$, 
where $F_i$ contains fermions and $B_i$ is made only from bosons:
\begin{align}
  F_1 &= -\frac{i}{8}gf_{abc}\phi'^a\eta^b\eta^c , &
  F_2 &= i\chi^a(D_0\psi_1^a - D_1\psi_0^a)\ , &
  F_3 &= \frac{i}{2}\psi_\mu^a D_\mu\eta^a
      + \frac{i}{2}g f_{abc}\phibar'^a \psi_\mu^b \psi_\mu^c \ ,  \nonumber\\
  B_1 &= -\frac{1}{8}g^2\left(f_{abc}\phi'^b\phibar'^c\right)^2 , &
  B_2 &= 2(F_{01}^a)^2 , &
  B_3 &=\frac{1}{2}D_\mu \phi'^a D_\mu\phibar'^a.
\end{align}
In the continuum, the kinetic terms and Yukawa interactions of the twisted
fermions are contained in the Dirac operator. Therefore, to find an appropriate
discretization of these terms, we simply replace the continuum Dirac operator
with the Wilson-Dirac operator. For example, we replace
\begin{equation}
 i\psi_0 D_0 \frac{\eta}{2} \Rightarrow
 \psi_0 \left((T^{-1})^T(CD_{\rm W})T^{-1}\right)_{\psi_0\eta} \frac{\eta}{2}.
\end{equation}
All dimensionful observables are measured in a unit of the 
dimensionful coupling $g$.

First, we present the result of the $\mathcal{O}_1$ case.
Figure~\ref{fig:o1-BF} shows that each of the bosonic and fermionic parts is
divergent in the continuum limit. The sum $\langle Q\mathcal{O}_1\rangle$ stays
finite after the reweighting by the Pfaffian, while the quenched result
diverges (Fig.~\ref{fig:o1}). In this theory, the cancellation of divergences in
$\langle Q\mathcal{O}_1\rangle$ is achieved by a balance between bosons' and
fermions' degrees of freedom. Our reweighted result is consistent with this
fact and the effect of dynamical fermions appears to be properly included by
the reweighting. As already noted, even in the supersymmetric continuum limit,
$\langle Q\mathcal{O}_1\rangle$ can be non-zero due to a finite
renormalization. The $\mathcal{C}$
dependence of $\langle Q\mathcal{O}_1\rangle$ is summarized in
Figure~\ref{fig:o1-Cdep}. To estimate the effect of the finite renormalization,
we determined the values of $\mathcal{C}$ which provide
$\langle Q\mathcal{O}_1\rangle=0$. They are $\mathcal{C}=1.047(51)$
at~$\beta=8$ and $\mathcal{C}=1.006(77)$ at~$\beta=20$. Almost no
$\beta$-dependence is observed. These values are significantly different from
$\mathcal{C}=0.65948255$ calculated in the continuum
limit~\cite{Suzuki-Taniguchi} and suggest that the effect of the finite
renormalization is certainly not negligible.

The cancellation of divergences is also realized both in $Q\mathcal{O}_2$ and
$Q\mathcal{O}_3$ cases (Fig.~\ref{fig:o2o3}). The $\mathcal{C}$-dependence,
however, is not manifest. Note that the plots have much more errors than that
of $\langle Q\mathcal{O}_1\rangle$ and this would imply that the
$\mathcal{C}$-dependence is smeared.
The difference between behavior of $Q\mathcal{O}_1$ and that of 
$Q\mathcal{O}_2$ and $Q\mathcal{O}_3$ could be accounted as a result of the
difference of the divergence of each operators. $B_1$ and $F_1$ have
logarithmic divergences, while $B_2$, $B_3$, $F_2$ and $F_3$ have quadratic
divergences.

%\begin{figure}
\FIGURE{
 \vspace{-.2em}

 \hfil
 \begin{minipage}{.48\linewidth}
  \includegraphics[angle=-90,width=\linewidth]{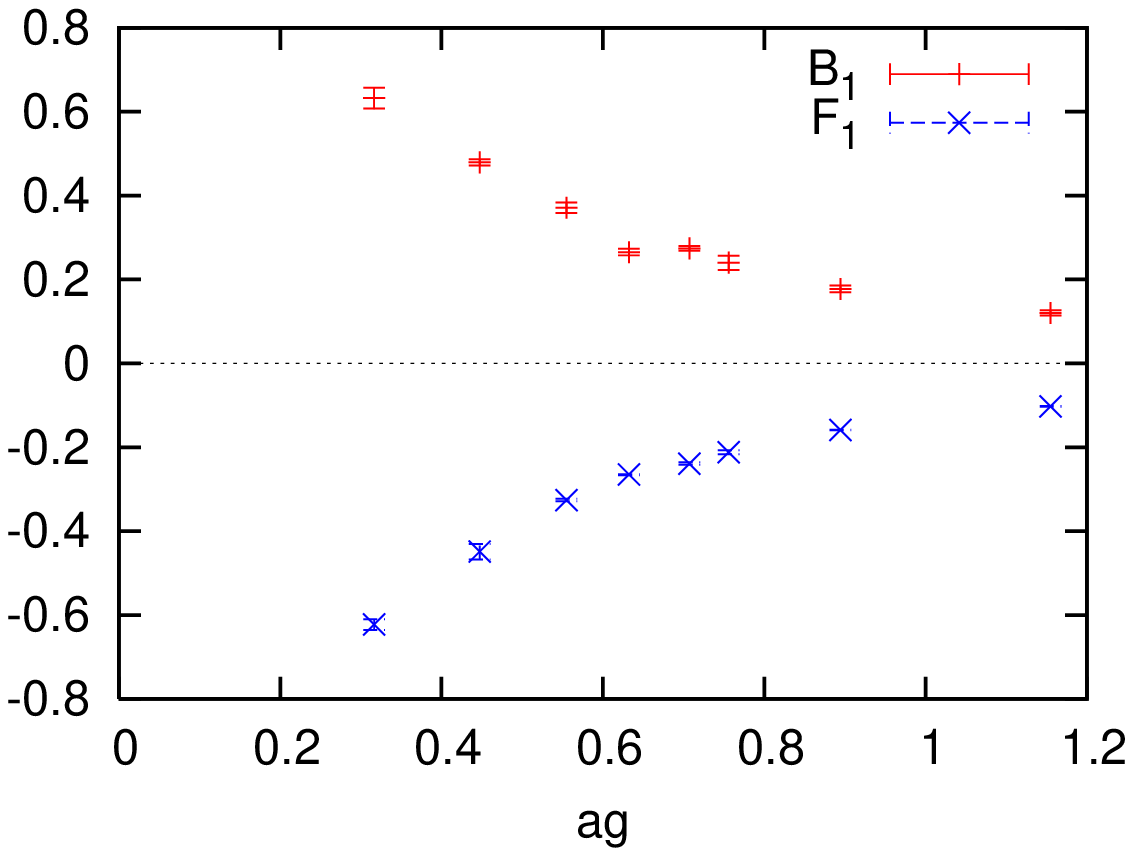}
 \end{minipage}
 \hfil
 \begin{minipage}{.48\linewidth}
    \includegraphics[angle=-90,width=\linewidth]{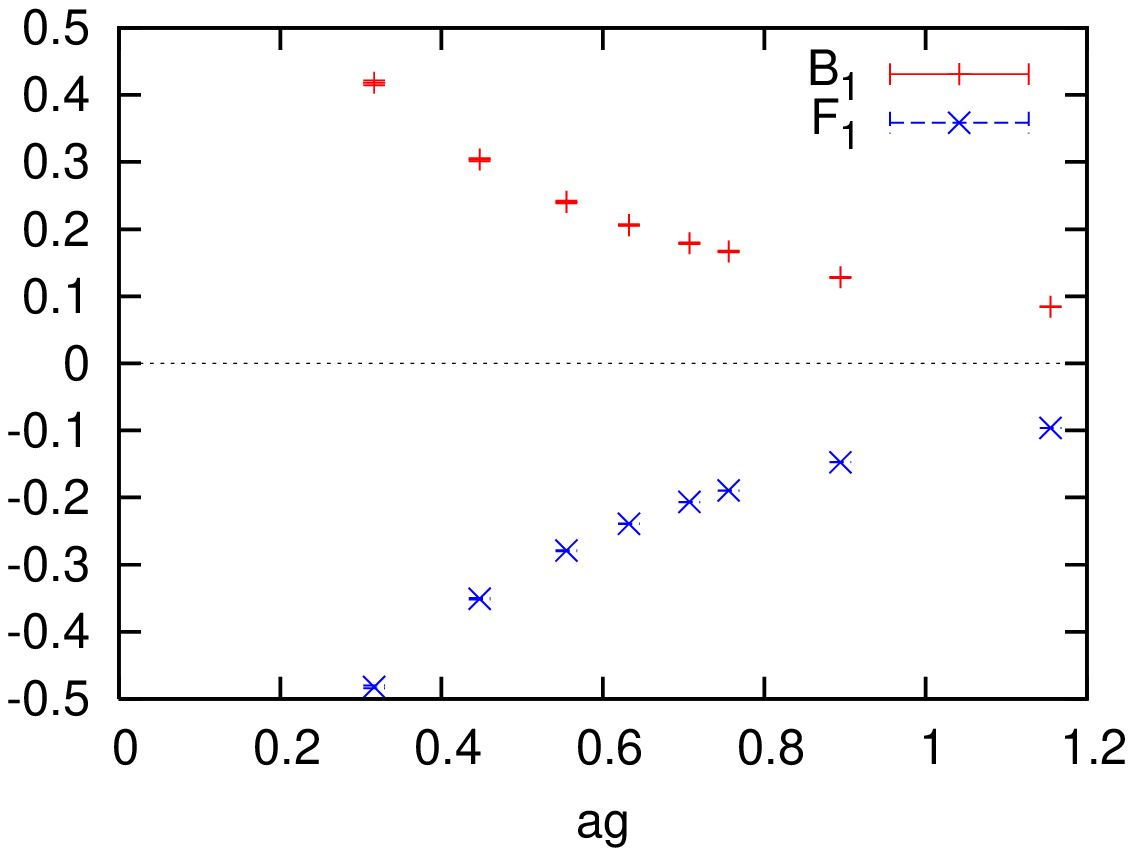}
 \end{minipage}
 \vspace{-.2em}

 \caption{The $a\to 0$ limit of $B_1$ and $F_1$, the left is reweighted and the right is quenched.
 The counter term is $\mathcal{C}=0.65948255$.
 The lattice size is $8 \times 8$.}
 \label{fig:o1-BF}
}
%\end{figure}

%\begin{figure}
\FIGURE{
 \vspace{-.2em}

 \hfil
 \begin{minipage}{.48\linewidth}
  \includegraphics[angle=-90,width=\linewidth]{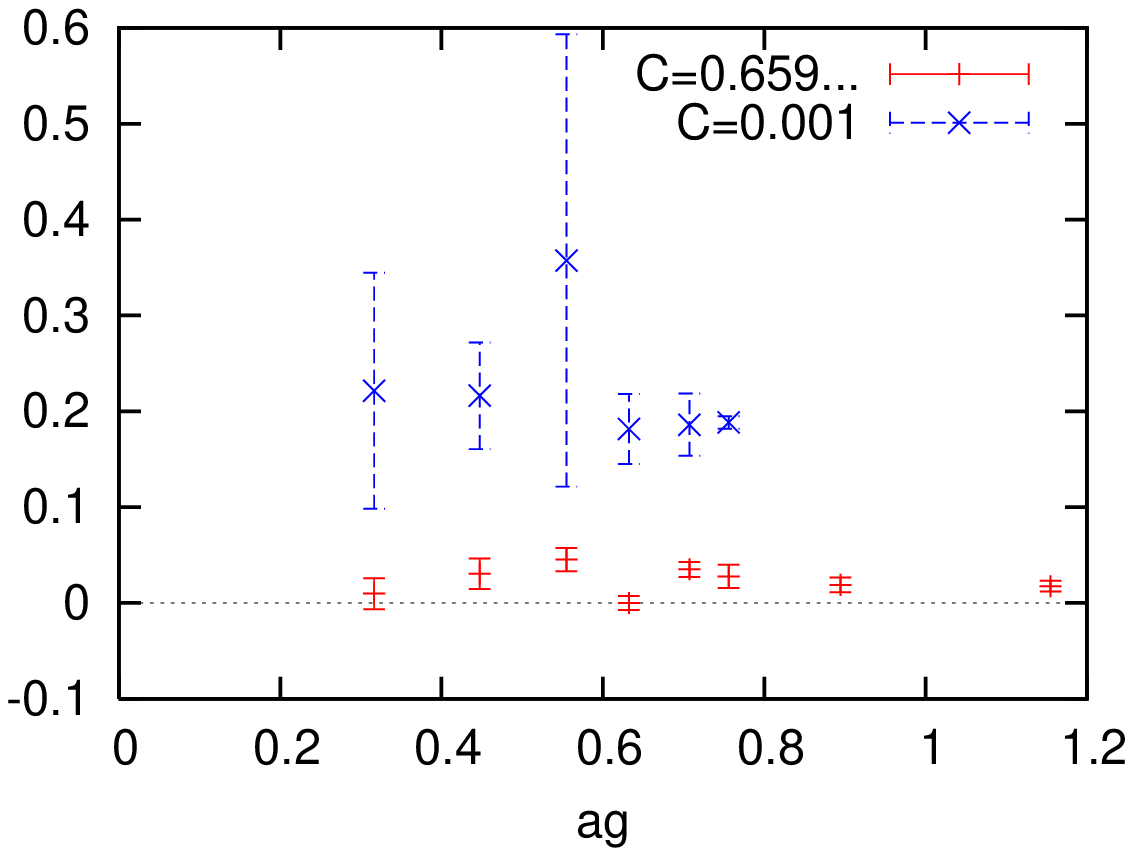}
 \end{minipage}
 \hfil
 \begin{minipage}{.48\linewidth}
    \includegraphics[angle=-90,width=\linewidth]{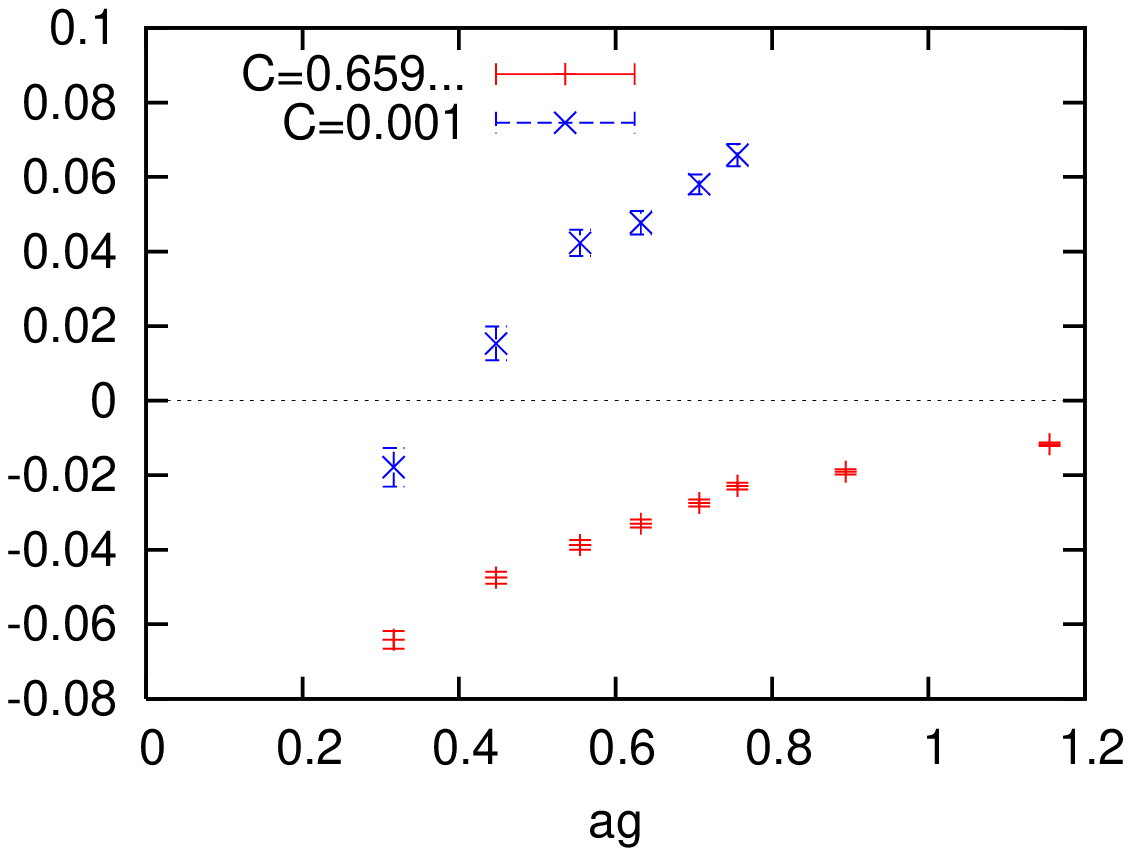}
 \end{minipage}
 \vspace{-.2em}

 \caption{The $a\to 0$ limit of  $\langle Q O_1\rangle$, 
 the left is reweighted and the right is quenched.
 The lattice size is $8\times 8$.}
 \label{fig:o1}
}
%\end{figure}

%\begin{figure}
\FIGURE{
 \vspace{-.2em}

 \hfil
 \begin{minipage}{.48\linewidth}
  \includegraphics[angle=-90,width=\linewidth]{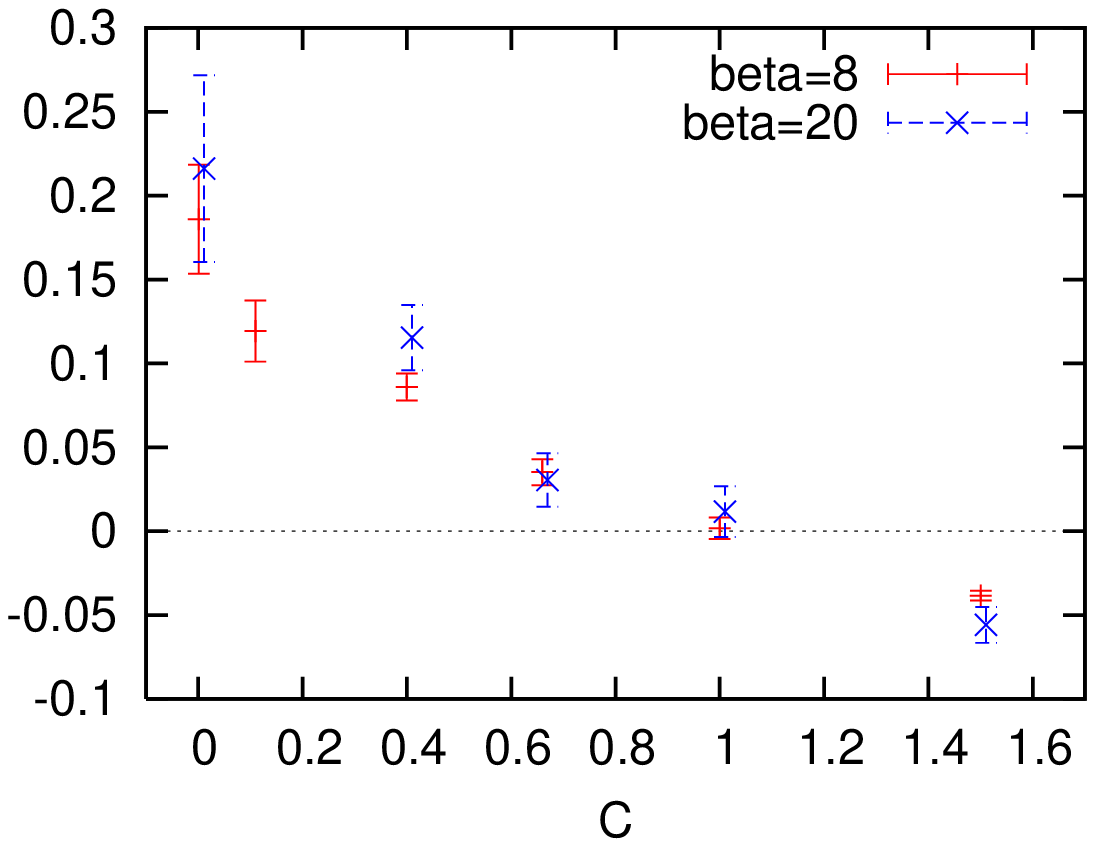}
 \end{minipage}
 \hfil
 \begin{minipage}{.48\linewidth}
    \includegraphics[angle=-90,width=\linewidth]{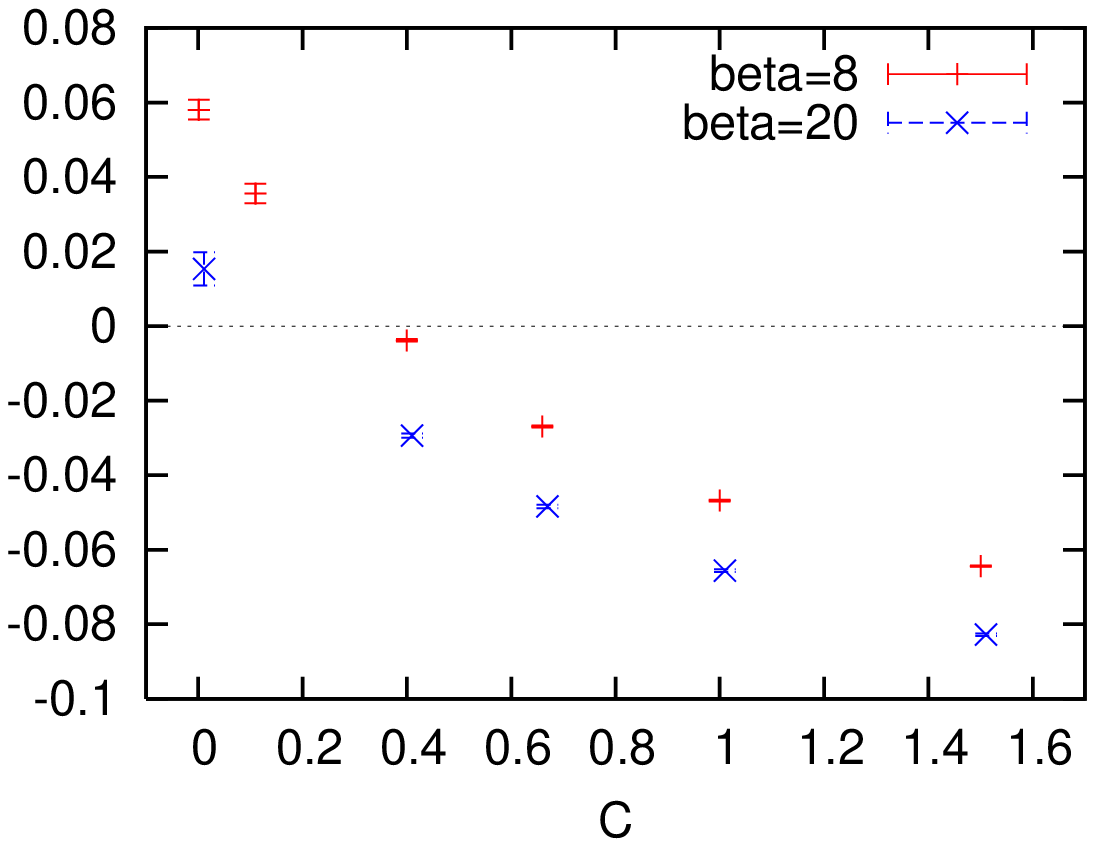}
 \end{minipage}
 \vspace{-.2em}

 \caption{The $\mathcal{C}$ dependence of $\langle Q O_1\rangle$, 
 the left is reweighted and the right is quenched.
 The lattice size is $8\times 8$.}
 \label{fig:o1-Cdep}
}
%\end{figure}

%\begin{figure}
\FIGURE{
 \vspace{-.2em}

 \hfil
 \begin{minipage}{.48\linewidth}
  \includegraphics[angle=-90,width=\linewidth]{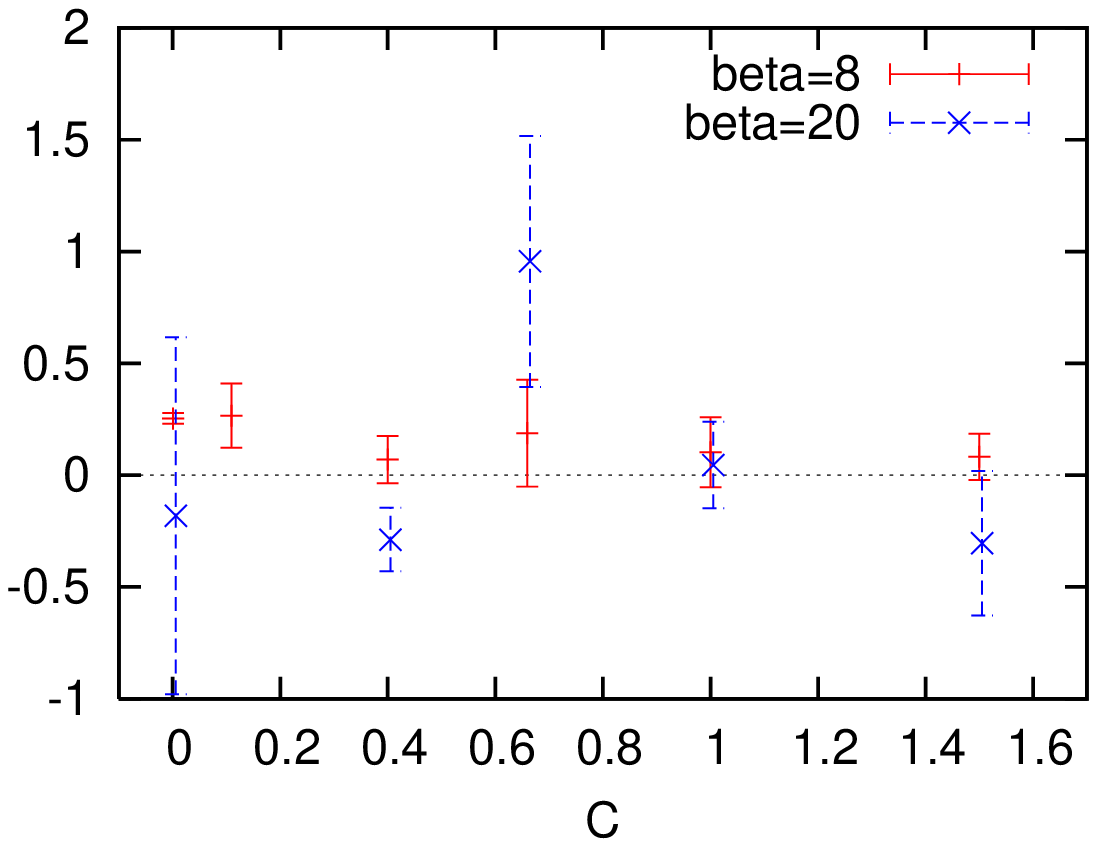}
 \end{minipage}
 \hfil
 \begin{minipage}{.48\linewidth}
    \includegraphics[angle=-90,width=\linewidth]{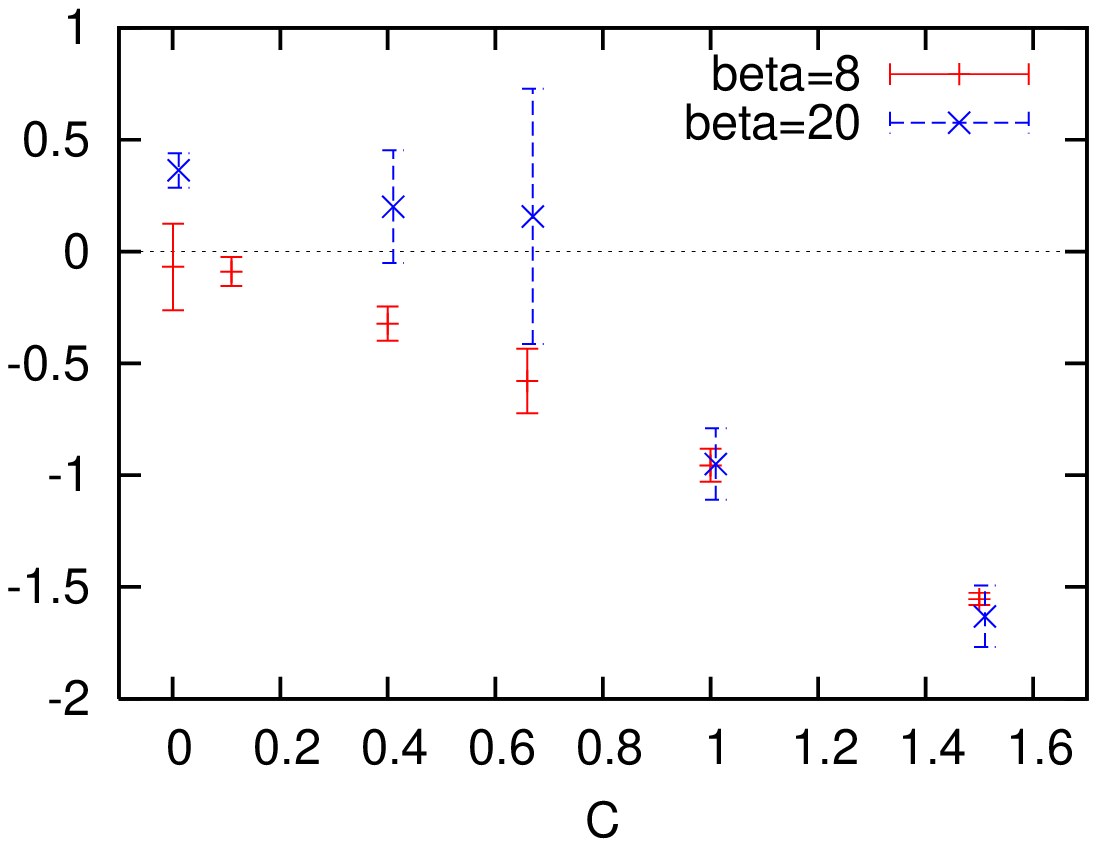}
 \end{minipage}
 \vspace{-.2em}

 \caption{The $\mathcal{C}$ dependence of $\langle Q O_2\rangle$ (left) and
 $\langle Q O_3\rangle$ (right). 
 These are reweighted and the lattice size is $8\times 8$.}
 \label{fig:o2o3}
}
%\end{figure}

\section{Two-point functions}
% WTi
Next we measure quantities with which the restoration of supersymmetry is
expected to be observed transparently. A supersymmetric Ward-Takahashi identity
indicates that  the following 2-point functions should have the same
functional form:\footnote{
The former function~$B$ exhibits a power law behavior in the continuum
theory~\cite{2dim-note}.
}
\begin{align}
 B &= 2i\left\langle j_{5\mu}(x) j_\nu(y)\right\rangle,  &
 F &= \left\langle
      \tr \left\{
       \gamma_\mu \gamma_5 (\phi^a + i\gamma_5\varphi^a)\psi^a(x)
       j_\nu^{\text{super}}(y)
       \right\}   
   \right\rangle, \label{eq:BandF}
\end{align}
where the bosonic currents are
\begin{align}
 j_\mu(x)
  &= \psibar^a \gamma_\mu \psi^a(x)\ , &
 j_{5\mu}(x)
  &= \psibar^a \gamma_\mu \gamma_5 \psi^a(x)
     + 2i \left\{ \phi^a \partial_\mu \varphi^a(x) -\varphi^a\partial_\mu\phi^a(x)\right\},
\end{align}
and the fermionic current is
\begin{align}
 j_\mu^{\text{super}}(x)
 &= \psibar^a\gamma_\mu \left\{
    \frac{1}{2}F^a_{\rho\sigma}\sigma_{\rho\sigma}
    -i\gamma_\rho D_\rho(\phi^a + i\gamma_5\varphi^a)
    -ig f_{abc}\varphi^b \phi^c \gamma_5
 \right\}(x)\ .
\end{align}
Figure \ref{fig:2pt-12x12} shows a typical result of the 2-point functions.
The number of the configurations we used is 101 and the bare fermion mass 
is $0$.
According to the scenario, we expect that a suitable choice of $\mathcal{C}$
should give the supersymmetric result, i.e., the identical spectra, while the
other choices of $\mathcal{C}$ should not. Unfortunately, errors in the plot
are too large to analyze the spectra although this is a result with the
quenched approximation. The point here is that we cannot distinguish the
difference of the counter term.
\FIGURE{
%\begin{figure}
 \hfil
 \begin{minipage}{.48\linewidth}
  \includegraphics[angle=-90,width=\linewidth]{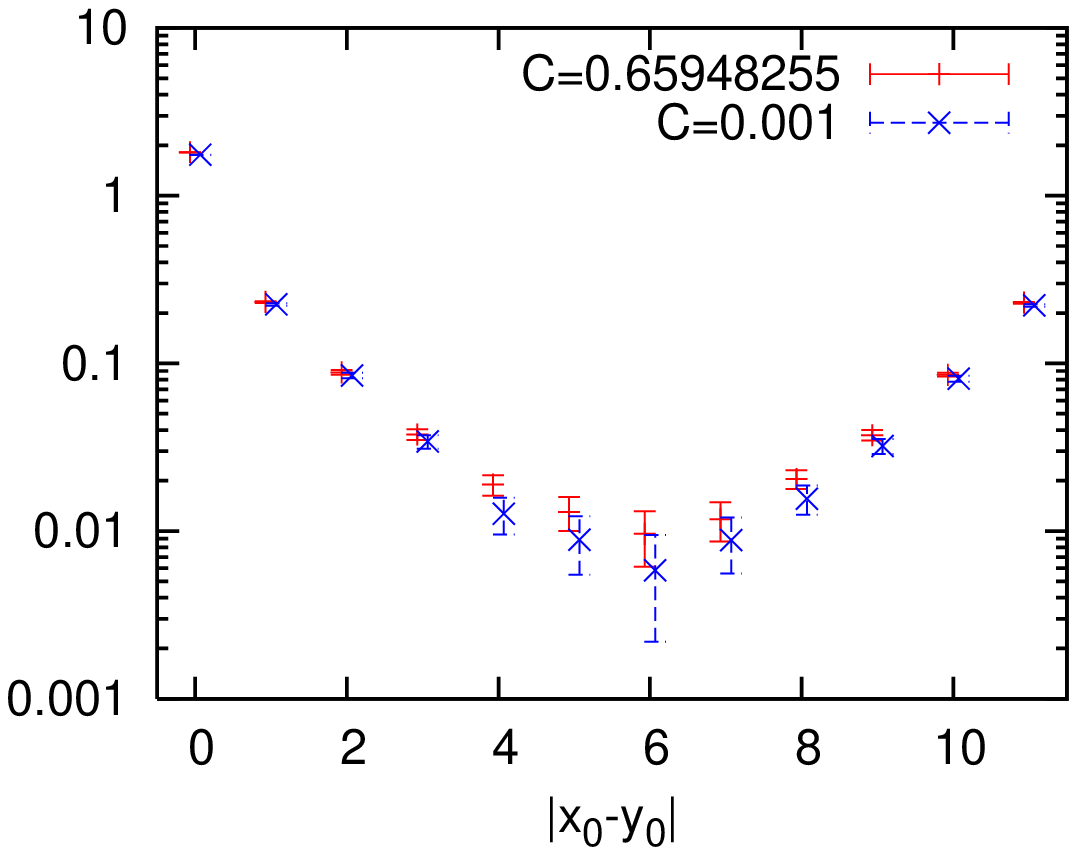}
 \end{minipage}
 \hfil
 \begin{minipage}{.48\linewidth}
    \includegraphics[angle=-90,width=\linewidth]{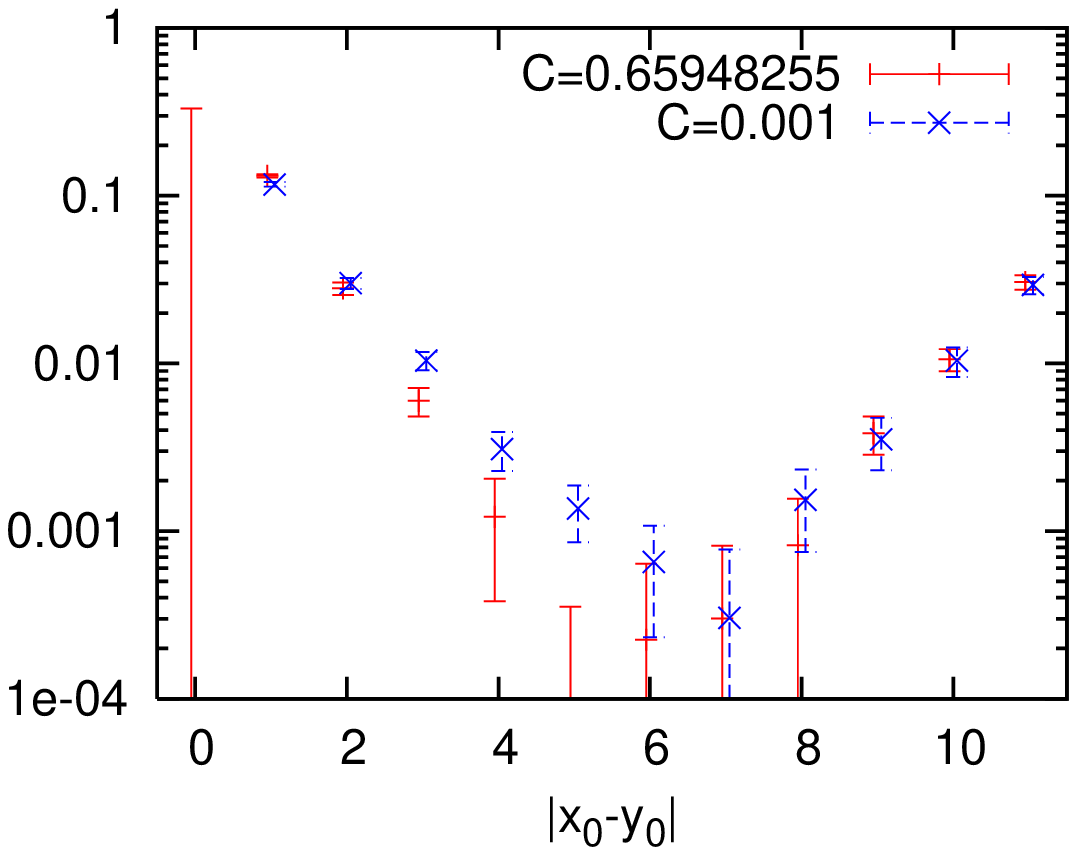}
 \end{minipage}
 \caption{The two-point functions, $B$ (left) and $F$ (right),
 in the quenched approximation.
 These are $\mu=1$, $\nu=0$ components of eq.~(\ref{eq:BandF}).
 The parameters are $\beta=18$ and $\mathcal{C}=0.65948255$, $0.001$ 
 on $12\times 12$ lattice.
 }
 \label{fig:2pt-12x12}
%\end{figure}
}

\section{Conclusion}

We observed 1-point functions and 2-point functions in a lattice formulation of
the two-dimensional $N=(2,2)$ super Yang-Mills theory. In our scenario, only 
the counter term coefficient $\mathcal{C}$ should be finely tuned. The 2-point
functions have large errors that we cannot compare the spectra associated with
a bosonic current and a fermionic current. The 1-point functions we observed
are finite in the continuum limit because of the fermion loop effect.
The result of a less divergent 1-point function depends on $\mathcal{C}$ and
is consistent with our scenario. To obtain the conclusive result from this
dependence, i.e., whether the scenario actually works or not, we need the
renormalization factor for the 1-point function.
Although the current result is not quite promising, we have some possible ways
to improve. A UV-filtered reweighting will help to reduce the errors after the
reweighting. The HMC algorithm with dynamical fermions is another option.
The result of 2-point functions suggests that the fermion or the scalar (or
both) is rather far from massless so that a negative bare mass of the fermion 
which reduces the physical mass may improve the sensitivity on the counter
term.

\acknowledgments
We would like to thank Yusuke Taniguchi for discussion at the early stage of
this work. We thank for computational resources of the RIKEN Super Combined
Cluster (RSCC). I.K. is supported by the Special Postdoctoral Researchers
Program at RIKEN. The work is supported in part by Grant-in-Aid for Scientific
Research, Nos.~18840045 (H.F.) and 18540305 (H.S.), and by JSPS and French
Ministry of Foreign Affairs under the Japan-France Integrated Action Program
(SAKURA).

\end{document}